\documentclass[12pt]{article}

\newcommand{\BV}{\left(\begin{array}{c}}
\newcommand{\EV}{\end{array}\right)}
\newcommand{\BM}{\left(\begin{array}{cc}}

\newcommand{\fep}{{\cal E}^0_0}
\newcommand{\rsqft}{<r^2>_{fit}}
\newcommand{\msqft}{<m^2>_{fit}}
\newcommand{\beq}{\begin{equation}}               
\newcommand{\eeq}{\end{equation}}                 
\newcommand{\bqry}{\begin{eqnarray}}              
\newcommand{\eqry}{\end{eqnarray}}                
\newcommand{\bqryn}{\begin{eqnarray*}}            
\newcommand{\eqryn}{\end{eqnarray*}}              
\newcommand{\NL}{\nonumber \\}                    
\begin{document}
\begin{titlepage}

\title{Tritium Beta Decay, Neutrino Mass Matrices \\
and Interactions Beyond the Standard Model }
\author{
G.J. Stephenson, Jr.\thanks{\small \em E-mail:
GJS@baryon.phys.unm.edu}\\
{\small \em Dept. of Physics \& Astronomy, University of New Mexico}\\ 
{\small \em Albuquerque, NM 87131 USA}\\
T. Goldman\thanks{\small \em E-mail:
tgoldman@lanl.gov}\\{\small \em 
Theoretical Division, MS B283, Los Alamos National Laboratory}\\ 
{\small \em Los Alamos, New Mexico 87545 USA}\\
B.H.J. McKellar\thanks{\small \em E-mail:
b.mckellar@physics.unimelb.edu.au}\\
{\small \em School of Physics, University of Melbourne}\\ 
{\small \em Parkville, Victoria 3052 Australia}
}

\date{June 1, 2000}
\maketitle
\vspace*{-5.1in}
\flushright{LA-UR-00-1327}
\vspace{-0.2in}
\flushright{UM-P-2000/013}
\vspace{-0.2in}
\flushright{hep-ph/0006095}\\
\vspace*{3.9in}
\begin{abstract}
The interference of charge changing interactions, weaker than the $V-A$
Standard Model (SM) interaction and having a different Lorentz
structure, with that SM interaction, can, in principle, produce effects
near the end point of the Tritium beta decay spectrum which are of a
different character from those produced by the purely kinematic effect
of neutrino mass expected in the simplest extension of the SM.   We
show that the existence of more than one mass eigenstate can lead to
interference effects at the end point that are stronger than those
occurring over the entire spectrum.  We discuss these effects both for
the special case of Dirac neutrinos and the more general case of
Majorana neutrinos and show that, for the present precision of the
experiments, one formula should suffice to express the interference
effects in all cases. Implications for "sterile" neutrinos are noted.
\end{abstract}
\flushleft
\end{titlepage}


\section{Introduction}

One of the outstanding problems of subatomic physics is that of
neutrino mass.  While this problem is being attacked on many fronts, the
most direct approach would seem to be the use of Tritium beta decay to
make a direct measurement of the mass of the electron anti-neutrino
~\cite{Zurich,LANL,LLNL,TRSK1,TRSK2,TRSK3,MNZ1,MNZ2,MNZ3}.  However, as
was known~\cite{KH} long before the advent of the Standard Model (SM), 
the exact form of the electron spectrum depends on the Lorentz
structure of the weak current involved in beta decay.  This observation
was recently revisited~\cite{SG},  wherein it was shown that possible
interference between the usual SM $SU(2)_W$ current and a (weaker)
non-SM current having a different Lorentz structure\footnote{Note that,
here and throughout this paper, we are looking at extensions of the SM
which are distinguishable at low energies.  This does not include
additional Left-chiral interactions~\cite{Marciano}.} can be
exacerbated by the presence of a CKM-like mixing~\cite{CKM} in the
lepton sector.  As we shall show in this paper, such effects could
produce a signal that might be interpreted as a negative value of the
square of the neutrino mass in an analysis which did not take
interference into account.  Whatever the eventual resolution of the
current experimental situation, our discussion makes clear the fact
that experimental analyses should not prejudice the result by assuming
solely a SM structure.

The concept of using a detailed measurement of the high energy portion
of the electron spectrum from nuclear beta decay to determine the mass
of the neutrino was introduced in Fermi's~\cite{Fermi} original paper.
In that work, he assumed that the interaction was due to a vector
current.  This gives rise (as we discuss below) to a different
dependence on neutrino mass than the $V-A$ current of the SM, as
evidenced qualitatively by the curves shown in his paper. As the
possibility of other Lorentz current structures was considered,
Koefed-Hansen~\cite{KH} pointed out the need to know the exact nature
of the currents to properly extract a neutrino mass.  Following the
establishment of the dominance of  $V-A$, Enz~\cite{Enz} recast the
problem in terms of possible interference between that current and
other (weaker) Lorentz currents and Jackson, Treiman and
Wyld~\cite{wyld} extended the discussion to include possible time
reversal invariance violation.

These earlier discussions were based on only one neutrino.  From LEP
data we now know that there is a net of three light neutrinos with full
SM coupling to the $Z^0$.  As we show below, both because the possible
interferences depend on the neutrino mass and because Tritium beta
decay experiments do not measure the neutrinos, it is appropriate to
frame the discussion in terms of mass eigenstates.  As this is more
easily visualized with Dirac neutrinos, we begin our discussion with
them, expanding to the more general and more widely accepted case of
Majorana neutrinos in a following section.  We then examine the effect
that any such interference will have on the extraction of the mass
parameters.

In a subsequent section, we present the effects of a possible
background scalar field, interacting only with
neutrinos~\cite{Clouds,MSWlike}, on the Tritium spectrum.  We also
examine the possibility that the interference effects could manifest
themselves at the upper end of the neutrino spectrum from other nuclear
beta decays~\cite{Maurice}.  Following this, we examine the
implications of the new interaction currents for neutrino neutral
current scattering, which devolve from the necessity of ``sterile''
neutrino interactions.  In the final section we reiterate our
conclusions and that experimentalists should use our functional form
for data analysis. We emphasize that this form must be used to avoid
introducing unwarranted theoretical prejudice into the interpretation
of the experimental results.


\section{Dirac neutrinos}

To present the general concepts that underlie our calculation in the
most accessible form, we first do the analysis for Dirac neutrinos.  In
later sections we shall deal with the more general case of Majorana
neutrinos.  By dealing with Dirac neutrinos, all of the manipulations
to calculate total rates from Feynman diagrams are standard and can be
found in any text on relativistic quantum mechanics.

\subsection{Mass eigenstates} \label{sec-masseig}

For the purpose of the discussion in this section, we assume the
existence of three massive Dirac neutrinos, $\nu^D_i $, with Dirac
masses $m^D_i$.  All three masses are less (really much less) than half
the mass of the $Z^0$.  We allow for the possibility that, as in the
quark sector, current eigenstates and mass eigenstates are not
necessarily identical.  In particular, we shall refer to the $\nu_f$,
$f = (e, \mu, \tau) $ as those linear combinations of mass eigenstates
which are coupled to the corresponding charged leptons by the $SU(2)_L
$ currents of the SM.  These current eigenstates are expressed in terms
of the mass eigenstates as
\[ \nu_f = \sum_i \cos \theta_f^i \nu_i \]
where the $\cos\theta_f^i$ are the direction cosines in the coordinate
system spanned  by the mass eigenstates.  (If two or more masses are
degenerate, we choose an orthogonal set.)  Under these assumptions, the
current eigenstates also form an orthogonal set.  This is depicted in
Fig.~1.

If there are additional objects, besides the known massive weak vector
bosons, that produce charge changing currents which couple to both the
quarks and the leptons, we may define neutral leptons which are current
eigenstates of this new interaction in exactly the same way, denoted by
$\hat{\nu}_f$, $f=(e, \mu, \tau)$, where now
\[ \hat{\nu}_f = \sum_i \cos \hat{\theta}_f^i \nu_i \]
and one would naturally expect that, in general, 
\[ \hat{\theta}_f^i \neq \theta_f^i . \]

\subsection{Interaction Hamiltonian}  \label{sec-intH}

Interactions beyond the SM must be weaker, at low energies, than the
usual Left-chiral SU(2) to avoid serious conflict with existing data.
Presumably this is due to the boson mediating the interaction being
much heavier than the known W's and Z, and/or the coupling constant
being smaller than that for the SM.  Absent a particular Grand Unified
Theory in which one wishes to embed the SM and the new interactions,
that is all one can say.

For low energy physics, like nuclear $\beta$-decay in general and
Tritium $\beta$-decay in particular, such new interactions can only
appear as effective currents in the four fermion formulation of the
theory with the usual space time structures of  $S, P, T, V$ or $A$.
Given the dominance of the SM, it is reasonable to recast this as
$S_{L}, S_{R}, T, R$ or $L$, where $R \sim (V+A)$ and $L \sim (V-A)$,
with a similar construction for $S_{L}$ and $S_{R}$ from $S$ and $P$.
The effect on the spectrum displayed below only occurs if there are
additional currents in the lepton sector that are different from $L$.
Corresponding changes in the hadron currents affect only the scale of
each new contribution.  To emphasize this fact, we employ an
unconventional notation; more conventional descriptions are given, for
example, by Enz~\cite{Enz}.

The effective low energy Hamiltonian for semi-leptonic decays is
\beq
H_I = \sum_{\alpha, \beta = S_{L},S_{R},R,L,T} G^{\alpha \beta}\sum_f 
\left( J^{\dagger}_{h\alpha}\cdot J_{f\beta} + h.c. \right)
\eeq
where, for example,
\beq
J_{f\alpha} =\overline{\psi_f} \Gamma_{\alpha} \psi_{\nu_f} \label{eq:cur}
\eeq
with $\psi_f$ representing a charged lepton of a given flavor,
$\psi_{\nu_f}$ a neutral lepton associated with that charged lepton
through the particular interaction (see the discussion below for more
detail). A similar construction can be made on the hadron side. Note
that, in this convention, the first Greek index in Eq.(1) refers to the
hadron current.  Explicitly,
\bqry
\Gamma_{S_L} &=& (1 - \gamma^5)  \nonumber\\
\Gamma_{S_R} &=& (1 + \gamma^5)   \nonumber\\
\Gamma_R &=& \gamma^{\mu}(1 + \gamma^5)
\nonumber \\
\Gamma_L &=& \gamma^{\mu}(1 - \gamma^5)
\nonumber \\
\Gamma_T &=& [\gamma^{\mu},\gamma^{\nu}]/2.
\eqry

Most off diagonal combinations ($\alpha \neq \beta $) in the sum
vanish, the exceptions being for $(S_R,S_L)$ and $(R,L)$.  For nuclear
beta decay in the SM, only $\beta = L$ and $\alpha = L,R$ survive.  The
relation to the basic parameters of the SM is given by~\cite{alphaW}
\beq
G^{RL} + G^{LL} = V_{ud}\frac{\pi\alpha_W}{\sqrt{2} M_W^2},
\eeq
where $V_{ud}$ is the appropriate element of the
hadronic CKM matrix~\cite{CKM},  $\alpha_W$ is
the fine structure constant for the $SU(2)_W$ of the SM,
$\alpha_W = \frac{g_W^2}{4 \pi } $,
where $g_W$ is the coupling constant,
and $M_W$ is the mass of the usual $W^{\pm}$.  To
complete the connection with conventional notation,
note that
\beq
(G^{RL} - G^{LL}) = \frac{G_A}{G_V} (G^{RL} + G^{LL}).
\eeq

The reason that the SM $SU(2)_L$ induces an hadronic Right-chiral
coupling is that the quarks are confined and their wave functions cause
$G_A$ to be renormalized with respect to $G_V$.   If there is an
$SU(2)_R$, mediated by a heavier vector boson or having weaker coupling
constant, or both, one still expects the hadronic current to be
modified.  However, in parallel to the case of the SM interaction, the
leptonic currents would be expected to be pure $V+A$.  In fact, one
also expects hadronic renormalization effects for $S$, $P$ and $T$.  In
addition, there will be, in general, a separate ``CKM'' matrix in the
quark sector for each interaction, mirroring the discussion of mass and
current eigenstates for neutrinos given above.

The various currents in the effective interaction Hamiltonian may be
generated by several different assumptions about physics beyond the
SM.  The most prominent are the exchange of a charge-changing scalar,
arising, for example, in supersymmetric models~\cite{susy}, or the
existence of a vector boson coupled to Right-chiral
fermions~\cite{lrsym} which may or may not mix with the W bosons of the
SM.  These are shown in Fig.~2, where we have included those cases
which may impact Tritium beta decay.  Although antisymmetric tensor
interactions have been proposed in some (mostly experimental)
contexts~\cite{Depom}, they would normally be expected to arise in
higher-loop order or by Fierz transformation from scalar leptoquark
interactions~\cite{GengLee} and so be exceptionally weak. In
particular, Voloshin~\cite{MBV} has obtained a very stringent bound on
the strength of  such a tensor interaction ($10^{-4} G_{F}$), which
applies unless a precise orthogonality holds for electron neutrino
states defined by the different interactions.

As we will show below, it will be difficult to distinguish amongst the
effective currents.  Given an effect in Tritium from some such
current, it will be even harder to trace that effect to a particular
diagram in Fig.~2.  However, this does mean that it makes sense to
pursue such effects in Tritium beta decay whatever one's prejudice
about a particular source of ``new physics''.

In terms of the diagrams of Fig.~2, we may define a new fine structure
constant, $\hat{\alpha}$ (to go with b) or the first diagram of c)) as
$ \hat{\alpha} = \frac{\hat{g}^2}{4 \pi }$.  Then the ratio of the
effective strength of this interaction at low energy to that of the SM
$SU(2)_L$ is given by $\hat{\rho}_X = \frac{\hat{g}^2}{g^2}
\frac{M_W^2} {{M}_X^2}$, where ${M}_X$ is the mass of the non-SM boson
being exchanged.

\subsection{Interference terms}  \label{sec-inter}

If there are additional bosons that couple to the Left-chiral Vector
fermion current~\cite{Marciano}, they can only renormalize $L$ and
cannot lead to any new structure near the end point. For this study
of Tritium beta decay all hadronic currents are evaluated in the
approximation of no nuclear recoil, which eliminates a pure
Pseudo-scalar current.

Nontheless, we allow for an arbitrary mixture of Scalar and
Pseudo-scalar, expressed as Left- and Right-chiral Scalar currents.
Both couple hadronically through S and couple independently to the
lepton current.  Consequently, we need to evaluate the effect on the
Tritium beta spectrum of possible interferences between the SM
Left-chiral Vector current and a Right-chiral Vector current, a
Right-chiral Scalar current and a Left-chiral Scalar current.  In this
work, a Left-chiral Scalar current is defined as that Scalar current
for which the Left-chiral projector acts on the neutrino field for the
current defined as in Eq.(\ref{eq:cur}), and a corresponding definition
for the Right-chiral case.

Since the hadronic currents must be renormalized and the fitting
procedure for any Tritium beta decay experiment fits the overall rate,
we only quote here the relative factors from the lepton traces. In
doing so, we make use of the fact that, when contracted with the hadron
traces and integrated over the outgoing neutrino directions, the only
terms that survive are those proportional to factors of the hadron mass
or energy, which differ negligibly due to the small size of the
momentum transferred to the final state hadrons (``no recoil''
approximation):
\[
\begin{array}{lr}
LL (SM)  & E_{\nu} E_{\beta} \\
RR           & E_{\nu} E_{\beta} \\
LR + RL           & - 2 m_{\nu} m_e \\
S_{L}S_{L}            & E_{\nu} E_{\beta}  \\
S_{R}S_{R}            & E_{\nu} E_{\beta}  \\
S_{L}S_{R} + S_{R}S_{L}           & -2 m_{\nu} m_e  \\
TT             & [E_{\nu} E_{\beta} - m_{\nu} m_e] \\
LS_{R} + S_{R}L      & - 2 m_{\nu} E_{\beta} \\
LS_{L} + S_{L}L      & 2 E_{\nu} m_e \\
LT + TL	& 2 [E_{\nu} m_e - m_{\nu} E_{\beta}]
\end{array}
\]
The negative sign of the terms with a factor of $m_{\nu}$ arises from
the fact that the neutral lepton in Tritium beta decay is an
anti-neutrino. 

For the interactions involving L (and by symmetry, R) currents, the
Lorentz inner product of the hadron trace with the lepton trace
produces an overall factor of $(G_{V}^{2} + 3 G_{A}^{2})$, and this
same factor occurs for the interference terms between the $W_{L}$ and
$W_{R}$ exchanges.

The case of of $W_{L}-W_{R}$ mixing is more complicated. It must
include a factor like this for the interference term with the SM in
which the $W_{R}$ couples directly to both leptons and hadrons as
above, but is more complicated and also involves other combinations,
such as $(G_{V}^{2} - 3 G_{A}^{2})$ for the term involving the
interference between the SM and the amplitude in which the mixed
propagator couples via $W_{L}$ to the hadrons and $W_{R}$ to the
leptons.

The corresponding result for the interference terms of a scalar current
with the SM currents is proportional to $G_{V}^{2}$ alone. This means
that the scalar interactions are somewhat less efficient at producing
new effects for the same relative strength as Right-chiral
interactions. 

Note that all interference effects of tensor interactions with the SM
(which effects are proportional to factors of $3 G_{A}^{2}$) are
encompassed by the scalar terms and so we will not explicitly discuss
tensor contributions further in this paper.

For our main purpose here, we need only consider the possibilities of
interference between the SM current and either a Right-chiral Vector
current (for Dirac neutrinos) or a Right-chiral Scalar current for
either Dirac or Majorana neutrinos. (The effect of interference with
the Left-chiral Scalar current appears in Sec.\ref{sec-neutep}, and an
additional tensor interaction is just a linear combination of the
Left-chiral Scalar and Right-chiral Scalar terms.)

\subsection{Pre-existing limits}  \label{sec-limits}

The best limits on the relative strengths of additional currents often
make use of interference effects~\cite{Herczeg}, and are deduced
assuming that the same neutrino is produced with the electron in all
cases.  As we discussed above, this is not a necessarily valid
assumption.  In fact, the interference effects need to be examined for
each mass eigenstate and, depending upon the details of the
interaction, could vanish over most of the spectrum even if the
effective coupling does not.

As an example, consider the possible effect on the spectrum due to the
interference between a Left-chiral Scalar and the SM Left-chiral
interaction.  As shown in the previous section, for each mass
eigenstate $\nu_i$, this interference is proportional to
$\hat{\rho}_{S_L} E_{\nu} m_e \cos \theta^{i}_{e} \cos
\hat{\theta^{i}_{e}}$.  For most applications, the neutrino masses are
negligible compared to $E_{\nu}$, so $q_{\nu} \cong E_{\nu}$ and
thresholds may be ignored.  This means that, when one sums over the
mass eigenstates, the result is proportional to $\hat{\rho}_{S_L}
\times \cos (x)$ where $x$ is the angle between $\hat{\nu_e}$ and
$\nu_e$, not simply to $\hat{\rho}_{S_L}$.

Recent limits~\cite{Herczeg} are given as
\begin{eqnarray*}
\hat{\rho}_R & \leq & .07 \\
\hat{\rho}_{S_R} & \leq & .1 \\
\hat{\rho}_{S_L} & \leq & .01 
\end{eqnarray*}
where the analysis has implicitly assumed unit value for $\cos (x)$.

\section{Majorana neutrinos}

In this section we discuss the more general possibility, arising from
the lack of any conserved charge for the neutrinos (at least at the low
energy scales below one-half the mass of the $Z_0$) that the mass
eigenstates correspond to Majorana neutrinos.

\subsection{Massive Weyl neutrinos}

We follow the development presented by Ramond~\cite{Ramond}.  The basic
object is a two complex component Weyl spinor which can be represented
under the Lorentz group in either of two inequivalent irreducible
representations, labelled conventionally as $(1/2 , 0)$ and $(0 ,
1/2)$.  These two representations of the same Weyl field are
isomorphic; the isomorphism connecting them is the operation of Charge
conjugation.  The usual Dirac spinor in the Weyl (or chiral)
representation~\cite{Itz} is constructed from one Weyl spinor in the
$(1/2 , 0)$ representation and an independent Weyl spinor in the $(0 ,
1/2)$ representation having the correct charges under appropriate
interactions.  A (Dirac) mass term coupling such representations allows
for a common global or local phase and, in particular, allows for a
Noether current that guarantees the conservation of particle number.

In this same representation, the four component representation of a
Majorana spinor is constructed by taking, for the Weyl spinor in the
$(0 , 1/2)$ representation, plus or minus times the Charge conjugate of
the Weyl spinor in the $(1/2, 0)$ representation.  This produces even
or odd Charge conjugation eigenstates, each of which contains exactly
the information contained in the original Weyl spinor.  The same
procedure as used for a Dirac mass term now produces a Majorana mass
term, only now, by construction, the two pieces of the spinor have
conjugate behavior under a global or a local phase and one cannot
construct a Noether current related to fermion number conservation.
Also, the interaction charges are opposite in the two parts of the
spinor, so this construction is not appropriate for any fermion with a
conserved charge.  Since both Left chiral $(1/2,0)$ and Right chiral
$(0,1/2)$ representations are present, this Majorana spinor is adequate
to describe all of the SM weak interactions and, if the mass is
non-zero, can mediate neutrinoless double beta decay.~\cite{HaxSt}

\subsection{Pure Weyl or ``see-saw'' neutrinos}

Appealing only to the SM without extensions, one need only consider
three Weyl neutrino fields associated with three (Majorana) mass
eigenstates, possibly rotated from the current eigenstates.  By
convention, these masses are labelled $m_L$.    With respect to Tritium
beta decay, there can be no interference with Right-chiral Vector
currents or with Right-chiral Scalar currents.  On the other hand,
interference with the Left-chiral Scalar current is possible, but,
since that current is the most constrained by other data, the
possibility of describing the data is less viable.

There is another nagging problem with pure, massive Weyl neutrinos.
This has to do with the possibility, inherent in the SM, that the
$SU(2)_L$ symmetry could be restored at high temperature.  In this
scenario, the SM Higgs loses its vacuum expectation value ($vev$), all
four weak bosons become massless and the weak charge is conserved.  The
last statement is contradicted by the existence of the Majorana mass,
which is certainly not connected to the standard Higgs.

To achieve such a mass, one generally introduces a new scalar field,
the Majoron~\cite{GR}, the sole purpose of which is generate a Majorana
neutrino mass via the $vev$ of the Majoron field.  As this $vev$ must
be small to avoid distortion of the ratio of $W$ and $Z$ masses beyond
experimental constraints, it may also be expected to evaporate as well
when the $SU(2)_L$ symmetry is restored, although this is not
guaranteed.

At some level, this is addressed by the usual see-saw~\cite{GMRS}.  In
that case one builds two distinct Majorana neutrinos, one from the Weyl
field described above and one from a Weyl field that could be used with
the first to build Dirac neutrinos.  One then assumes that the usual
Higgs produces some Dirac masses ($m_D$) (presumably of the order of
charged lepton masses, although that is not critical) and that the new
Weyl field, which is sterile to all known interactions, develops a
Majorana mass through some means that does not affect the SM.  That
mass is conventionally termed $M_R$, and is actually a $3 \times 3$ (or
larger, depending on the full model) matrix in flavor space.

This has the well known pleasant consequence that the masses of the
light (active) neutrinos are of the order of $\frac{m_D^2}{M_R}$,
leading to very small neutrino masses.  However, it should clearly be
noted that there is no principle leading to this construction.  There
is also no guarantee that the rank of $M_R$ is three, so there could be
vastly  different patterns~\cite{miami} in different generations.  In
this description (assuming rank three), the three light neutrinos are
essentially indistinguishable from the pure Weyl case insofar as
Tritium beta decay is concerned.  However, since $m_D$ is assumed to be
proportional to the Higgs vacuum expectation value, the problem with
symmetry restoration is avoided.  We do note, however, that $L-R$
symmetric models with a (relatively) low lying Right-chiral scale face
the same problem of symmetry restoration.

\subsection{Pseudo-Dirac neutrinos}

For many years there has been discussion of the possibilities
associated with a very small Majorana mass, either compared with Dirac
masses or masses only off-diagonal in flavor space, the so-called
``Pseudo-Dirac'' case~\cite{WolfpD,SchV,BilPet}.  In this case, which
has recently been revisited in the literature~\cite{GSMK,KC}, for
Tritium beta decay there is really no distinction from the pure Dirac
case.  Any time evolution will be so slow that the implications for the
spectra considered in this paper are negligible.

\section{Implications for Tritium beta decay}

We now examine the effect of the possibilities we have discussed on the
analysis of the beta spectra in Tritium beta decay.  We recognize that
no attempt should be made to extract any parameters from the published
literature, as proper analysis requires inclusion of specific
experimental details.  What we shall show is that, for parameter ranges
which are not in conflict with other experimental results, the
combination of non-vanishing neutrino masses and an interference at
very low neutrino energy can cause an analysis, which assumes no
interference, to produce negative values of $m_{\nu}^2$.  Furthermore,
there is a dependence of the extracted value on the range of $\beta$
energies used for the fit (for differential spectra) or the $\beta$
bias energy (for integral spectra) which dependence matches that
reported by some experiments~\cite {LANL,MNZ1,MNZ2,MNZ3}.

\subsection{Differential spectra}

Our discussion in this paper is directed at those experiments using
molecular Tritium as a source.  Other sources require well known
modifications to the discussion which are no different for our case
than for the SM.

For the SM only, the differential spectrum, $\frac{dN}{dE_{\beta}}$, is
already a sum of several individual spectra for each possible end point
corresponding to a particular final energy of the $^{3}He-T$ molecular
ionic system.  Strictly speaking, the sum should include an integral
over the continuum of ionic breakup states, although the analysis may
find that it is adequate to represent that continuum with a finite
sum.  This may be expressed as
\beq
\left(\frac{dN}{dE_{\beta}} \right)_{SM} =
\sum_i P_i \left(\frac{dN}{dE_{\beta}} ({\cal E}^i_0)
\right)_{SM} \label{eq:desm1}
\eeq
where $P_i$ is the probability of leaving the ionic system in the
$i^{th}$ final state and ${\cal E}^i_0$ is the maximum energy available
to the $\beta$ for that final state, assuming that $m_{\nu} = 0$.  At
present, the $P_i$ are calculated~\cite{SF} and current experiments
estimate that the theoretical uncertainty which this introduces
constitutes a small part of the error budget.  The end point of the
spectrum (for $m_{\nu} = 0$) is then $\fep$ where $i = 0$ denotes the
ground state of the molecular ion.  Of course, the expression in
Eq.(\ref{eq:desm1}) is theoretical only.  In fitting to a particular
experimental spectrum, any energy dependent experimental corrections
must either be included explicitly or allowed to vary, within reason,
in the fitting procedure.

In this paper we shall illustrate the effects of possible interferences
on the theoretical spectrum appropriate to $\fep$.  The extension to
the full spectrum follows Eq.(\ref{eq:desm1}).  For zero mass
neutrinos, where the SM weak current eigenstate may be taken as the
only neutrino produced, we have
\bqry
\left(\frac{dN}{dE_{\beta}}(\fep)\right)_{SM} & = &
K F(E_{\beta}) q_{\beta} q_{\nu} E_{\beta} E_{\nu} \nonumber \\
  & = & K F(E_{\beta}) q_{\beta} E_{\beta} (\fep - E_{\beta})^2 ,
\eqry
where $K$ includes all those quantities, such as the nuclear matrix
element, the source strength and the overall experimental efficiency,
subsumed by the measured normalization, $F(E_{\beta})$ is the Fermi
function, $q_{\beta}$ and $E_{\beta}$ are the momentum and relativistic
energy of the $\beta$ and $q_{\nu}$ and $E_{\nu}$ are the momentum and
relativistic energy of the antineutrino.  The second line reflects the
assumptions that $E_{\nu} = \fep - E_{\beta}$ and that $q_{\nu} =
E_{\nu}$ for a massless neutrino.

The simplest extension to the SM consists of assuming that there are no
additional interactions, that there is still only one relevant neutrino
(strictly, $\bar{\nu}_e$), but that this neutrino may have a mass.  In
this case we make the replacement $q_{\nu} = \sqrt{E_{\nu}^2 -
m_{\nu}^2}$ and fit
\beq
\left(\frac{dN}{dE_{\beta}} \right)_{1} =
\sum_i P_i \left(\frac{dN}{dE_{\beta}} ({\cal E}^i_0)
\right)_{1} \label{eq:desm2}
\eeq
where, for example,
\bqry
\left(\frac{dN}{dE_{\beta}}(\fep)\right)_{1} & = &
K F(E_{\beta}) q_{\beta} E_{\beta} E_{\nu}
\sqrt{E_{\nu}^2 - m_{\nu}^2} \Theta(\fep -E_{\beta} - m_{\nu}) \nonumber \\
  & = & K F(E_{\beta}) q_{\beta} E_{\beta} (\fep - E_{\beta})^2 
\sqrt{1 -\frac{m_{\nu}^2}{(\fep - E_{\beta})^2}} \nonumber \\
  &   & \times \Theta(\fep -E_{\beta} - m_{\nu}), \label{eq:desm3}
\eqry
treating $m_{\nu}^2$ as an additional parameter.  $\Theta(z)$ is the
Heavyside function.  As is well known, the reported values of
$m_{\nu}^2 < 0$ are an indication of an excess of counts near the
endpoint over the expectation for zero mass, with all other parameters
determined from the robust spectrum far below the end point.

The first obvious extension is to include the possibility of different
mass eigenstates, with the SM current eigenstates being mixtures of
those mass eigenstates denoted by $\nu_k$,
\beq
\bar{\nu}_e = \sum_k \cos \theta^k_e \bar{\nu}_k \label{eq:nubar}
\eeq
giving
\bqry
\left(\frac{dN}{dE_{\beta}}(\fep)\right)_{2} & = & K F(E_{\beta}) 
q_{\beta} E_{\beta} (\fep - E_{\beta})^2 \sum_{k} \cos^2 \theta^k_e 
\sqrt{1 -\frac{m_{k}^2}{(\fep - E_{\beta})^2}} \nonumber \\
&   & \times \Theta(\fep -E_{\beta} - m_{k}), \label{eq:desm4}
\eqry
which does not improve the fit near the endpoint~\cite{LLNL}.

The next extension, which is the point of this paper, is to include the
possibility of interference with additional interactions which appear,
at the low energies associated with nuclear beta decay, as currents
with a different character under Lorentz transformations.  As stated in
Sec.\ref{sec-inter}, we confine ourselves to Scalar currents, which can
play a role for any neutrinos, and Right-chiral Vector currents which
affect only Dirac and pseudo-Dirac neutrinos.

Such currents will produce direct effects on the rate, proportional to
$\hat{\rho}_X^2$ as well as interference terms proportional to
$\hat{\rho}_X$.  The former will be hard to observe and, for Tritium
beta decay, will be absorbed into normalization factors, but are
included here for completeness.

In subsections \ref{sec-masseig} and \ref{sec-intH} and in
Eq.(\ref{eq:nubar}) above, we used the notation $\cos\hat{\theta}_f^i$
to refer to the direction cosines for new current eigenstates coupled
to the charged lepton $f$ relative to the mass eigenstate labelled by
$i$.  In this case, the only flavor eigenstate considered is $e$ and we
want to consider different possible currents, so we change the notation
to read $\cos \theta_{iX}$ where $i$ refers to the mass eigenstate and
$X$ refers to the current $R$, $S_R$ or $S_L$ (and is omitted for the
SM current).  We further define $\rho_X = \hat{\rho}_X (ME)_X$ where
$(ME)_X$ accounts for the ratio of the hadronic matrix element of the
designated current relative to that of the SM in the particular nuclear
system, here $A=3$. This includes the (quark) CKM matrix elements
appropriate to that non-SM interaction.

At the risk of being overly pedantic, we present the discussion for
both of the cases under consideration first separately and then combine
them.  For a Right-chiral Vector current (without L-R mixing), we make
the substitution
\[
\cos^2 \theta_i E_{\beta} E_{\nu} \rightarrow \cos^2 \theta_i E_{\beta} 
E_{\nu} + \cos^2 \theta_{iR} \rho_R^2 E_{\beta} E_{\nu} - 
2 \cos \theta_i \cos \theta_{iR} \rho_R m_e m_{i}. \]
Defining \[ \epsilon_{iR} = \rho_R \frac{\cos \theta_{iR}}{\cos\theta_i} \]
we may recombine this as 
\[
\cos^2\theta_i E_{\beta} E_{\nu}  ( 1  - 2 \epsilon_{iR} 
\frac{m_e}{E_{\beta}} \frac{m_i}{E_{\nu}} +\epsilon_{iR}^2  ). \]

We remind the reader that the sign of the middle term arises from the
fact that the neutral lepton in Tritium beta decay, in terms of Dirac
fields, is an anti-neutrino.  Since $\epsilon_{iR}$ contains the ratio
of direction cosines, it also may have either sign.  This allows for
positive interference near the end point.  Similar remarks apply
throughout this section.

Turning now to the other case of interest to us here, of a Right-chiral
Scalar interaction (as defined in Secs.\ref{sec-intH},\ref{sec-inter}),
we make the substitution 
\[ 
\cos \theta_i^2 E_{\beta} E_{\nu} \rightarrow 
\cos \theta_i^2 E_{\beta} E_{\nu} + \left(
\frac{G_{V}^{2}}{G_{V}^{2} + 3 G_{A}^{2}} \right) \]
\[ \hspace*{2.in} \times \left[ \cos^2 \theta_{iS_{R}} \rho^2_{S_R} 
E_{\beta} E_{\nu} - 2 \cos\theta_i \cos\theta_{iS_{R}} \rho_{S_R} 
E_{\beta} m_{i} \right] 
\] 
And defining 
\[ 
\epsilon_{iS_{R}} = \rho_{S_R} \frac{\cos \theta_{iS_{R}}}{\cos\theta_i} 
 \]
we obtain the expression
\[ \cos^2\theta_i E_{\beta} E_{\nu} \left[ 1 + \left( 
\frac{G_{V}^{2}}{G_{V}^{2} + 3 G_{A}^{2}} \right) 
\left( - 2 \epsilon_{iS_{R}} \frac{m_i}{E_{\nu}} + 
\epsilon_{iS_{R}}^2 \right) \right]. \]

Finally, combining both of these possibilities (and keeping in mind that
any tensor interaction contributions are encompassed by scalar terms),
we obtain
\beq
\left(\frac{dN}{dE_{\beta}}\right) = \sum_i P_i \left(\frac{dN}{dE_{\beta}}({\cal E}^i_0)\right) \label{eq:enchila}
\eeq
with
\bqry
\left(\frac{dN}{dE_{\beta}}({\cal E}^i_0)\right) 
& = & KF(E_{\beta}) q_{\beta} ({\cal E}^i_0 - E_{\beta}) \sum_k 
\Theta({\cal E}^i_0 - E_{\beta} - m_k) \nonumber \\ 
&   & \times E_{\beta} ({\cal E}^i_0 - E_{\beta}) 
\sqrt{1-\frac{m_k^2}{({\cal E}^i_0-E_{\beta})^2}} \nonumber \\ 
&   & \times \left( [\cos^2 \theta_k + \cos^2 \theta_{kR} \rho_R^2 + 
\left( \frac{G_{V}^{2}}{G_{V}^{2} + 3 G_{A}^{2}} \right) \cos^2 
\theta_{kS_{R}} \rho_{S_R}^2] \right. \nonumber \\ 
&   & \left. - 2 m_e m_k [ \cos \theta_k \cos \theta_{kR} 
\rho_R ] \right. \nonumber \\ 
&   & \left. -2 m_k E_{\beta} \left( \frac{G_{V}^{2}}{G_{V}^{2} + 
3 G_{A}^{2}} \right) [\cos \theta_k \cos \theta_{kS_{R}} \rho_{S_R}] 
\right) .  \label{eq:finaldif}
\eqry

While this expression has everything in it, it is not particularly
useful for fitting experimental data.  We can cast it into a more
useful form by noticing that, near the end point, the dependence on
$E_{\beta}$ is very gentle and one is really interested in the
dependence on $E_{\nu} = ({\cal E}^i_0 - E_{\beta})$.  Furthermore, the
product $F(E_{\beta}) q_{\beta}$ is nearly constant for $\beta$
energies near the end point.  Note also that the ratio
$\frac{m_e}{E_{\beta}}$ varies by less than $2$ parts in $10^3$ as the
kinetic energy of the $\beta$ varies from $17.5 keV$ to $18.5 keV$, a
much wider range than is used in modern fits~\cite{TRSK3,MNZ3}.  These
observations suggest the following approximations.

First, absorb $F(E_{\beta}) q_{\beta} E_{\beta}$ into a new ``constant'' 
\[ K' = K F(E_{\beta}) q_{\beta} E_{\beta}. \]

Second, for some average value of $E_{\beta}$, take
$\frac{m_e}{E_{\beta}} = \frac{m_e}{<E_{\beta}>}$ as
constant\footnote{Since it is buried in a parameter to be fit, the
actual value is not important.} and define
\[
\varepsilon_k = \epsilon_{kR}^2 + \epsilon_{kS_{R}}^2 \left( 
\frac{G_{V}^{2}}{G_{V}^{2} + 3 G_{A}^{2}} \right) \]
and
\[ \phi_k = - 2\frac{m_k}{(1+\varepsilon_k)}\left[ \frac{m_e}{<E_{\beta}>} \epsilon_{kR} 
 + \epsilon_{kS_{R}} \left( \frac{G_{V}^{2}}{G_{V}^{2} + 3 G_{A}^{2}} 
\right) \right] . \]

\pagebreak
Then
\bqry
\frac{dN}{dE_{\beta}}({\cal E}^i_0) & \cong & K' \sum_k \cos^2 \theta_k 
({\cal E}^i_0 - E_{\beta})^2 (1 + \varepsilon_k ) \left[ 1 +
\frac{\phi_k}{({\cal E}^i_0 - E_{\beta})} \right] \nonumber \\
&   & \times \sqrt{1-\frac{m_k^2}{({\cal E}^i_0 - E_{\beta})^2}} 
\Theta({\cal E}^i_0 - E_{\beta} - m_k) \label{eq:apprdif}
\eqry

\subsection{Integral spectra}

While many of the experiments performed over the last few decades are
differential measurements~\cite{Zurich,LANL,LLNL}, the two ongoing
experiments are integral measurements which accept all $\beta$s with
energy above some cutoff 
energy~\cite{TRSK1,TRSK2,TRSK3,MNZ1,MNZ2,MNZ3}.  To obtain the
theoretical form for the expected count rate for a given set of
parameters, we should integrate Eq.(\ref{eq:finaldif}) over $E_{\beta}$
from $E_{\beta}^C$ to $\infty$.  This daunting prospect requires
numerical treatments.  However, the form given in Eq.(\ref{eq:apprdif})
is both a very good approximation and amenable to analytic
integration.  The integral needs to be evaluated for each endpoint
energy ${\cal E}^i_0$ and for each mass eigenvalue $m_k$.  Let us
explicate the procedure for $\fep$ and one mass, $m_k$.

We need to evaluate
\[  K' \cos^2 \theta_k \int_{E^C}^\infty dE_{\beta} 
({\cal E}^0_0 - E_{\beta})^2 \sqrt{1 -\frac{m_k^2}{({\cal E}^0_0 
- E_{\beta})^2}} \Theta({\cal E}^0_0 - E_{\beta} - m_k) \]
\[ \hspace*{2.in} \times (1+\varepsilon_k) 
\left[1+ \frac{\phi_k}{({\cal E}^0_0 - E_{\beta})} \right]. \]
Changing variables to $E_{\nu} = (\fep - E_{\beta})$ and defining
$E_{\nu}^C = (\fep - E_{\beta}^C)$, we get
\bqry
N(E_{\nu}^C) & = &
K' \cos^2 \theta_k \int_{m_k}^{E_{\nu}^C} dE_{\nu} 
\sqrt{ 1 - \frac{m_k^2}{E_{\nu}^2}}E_{\nu}^2(1 + \varepsilon_k ) \left(1+\frac{\phi_k}
{E_{\nu}} \right) \nonumber \\ 
& = & K' \cos^2 \theta_k (1+ \varepsilon_k ) \left[ \frac{1}{3} 
((E_{\nu}^C)^2 - m_k^2)^{3/2} \right. \nonumber \\ 
& &  \left. + \frac{\phi_k}{2} \left(E_{\nu}^C \sqrt{(E_{\nu}^C)^2 - m_k^2} 
- m_k^2 \ln{(\frac{E_{\nu}^C + \sqrt{(E_{\nu}^C)^2 - m_k^2}}{m_k})} 
\right) \right]. \label{eq:apprint}
\eqry

\subsection{Fitting fitted differential spectra}

We emphasize, in this subsection and the next, that we attempt neither
to fit data nor to extract reliable values of parameters, as that must
be done with full knowledge of all experimental details.  What we shall
do is treat the published values of $\msqft$ as a representation of the
data. We then ask what parameters will produce similar values of
$\msqft$ if a fit were to be made, using only the SM expression, to our
theoretical spectrum which includes interference effects.  The purpose
of doing that, in this paper, is to demonstrate that the interference
terms apparently do allow a representation of the experimental data.

Furthermore, to simplify both the discussion and our task, we assume
that the various experimental groups have correctly performed the
weighted sums over the final states of the molecular ionic system.
Therefore, we need only study the effects of interference compared with
the SM analysis on one branch, which we shall take to be the ground
state branch.  In fact, the entire analysis of these next two
subsections goes through unchanged for other branches, but the
equations become needlessly cumbersome for simply the demonstration of
our point.

Let us now further assume for simplicity that only one mass eigenstate
is important for the fit near the end point. Let that mass eigenvalue
be denoted as $m_1$.

Since, for Tritium beta decay, the range of $E_{\beta}$ is
\[
m_e \leq E_{\beta} \leq 1.035 m_e
\]
one will not be able to distinguish between interference terms
involving $E_{\beta}$ and $m_e$.  On the other hand, since we are
interested in effects where $E_{\nu}$ varies significantly compared to
$m_{\nu}$, those terms will have very different effects on the
spectrum.

Now scale $E_{\nu}$ and the $\phi_i$ in units of $m_1$,
\bqry
x & = & E_{\nu}/m_1 \nonumber \\
f_i & = & \phi_i /m_1
\eqry
In this case, defining $Y_1 = \cos^2\theta_1 (1 + \varepsilon_1)$, the
differential spectrum becomes
\beq
\frac{dN}{dE_{\beta}} = K' Y_1 x^2 (1 + \frac{f_1}{x})
\sqrt{1 - \frac{1}{x^2}} \Theta(x-1).  \label{eq:simp1}
\eeq
If we were to fit this with a formula for the spectrum which is derived
under the assumption that there is only one neutrino and that it has a
mass extracted from the spectrum as $<m^2>_{fit}$, we would fit
Eq.(\ref{eq:simp1}) with the function
\beq
\frac{dN}{dE_{\beta}} = K' Y_T x^2 \sqrt{1-\frac{<r^2>_{fit}}{x^2}}  
\label{eq:simp2}
\eeq
where $<m^2>_{fit} = m_1^2 <r^2>_{fit}$ is the extracted (apparent)
value of the neutrino mass-squared, and $Y_T = \sum_k \cos^2\theta_k (1
+ \varepsilon_k)$. The reason that $Y_T$ appears is that we assume that
the experimental normalization of the data is taken from a region of
lower electron energies where neutrino mass effects are presumed to be
negligible and all mass eigenstates (and currents) contribute.

Setting Eqs.(\ref{eq:simp1}) and (\ref{eq:simp2}) equal, at some
particular value of $x$, gives an equation for $<r^2>_{fit}$,
\bqry
<r^2>_{fit} &=& x^2[1-\zeta^{4}] - 2 x f_1 \zeta^{4} \nonumber\\
           &  & + [1 - f_1^2] \zeta^{4} + 2 x^{-1} f_1 \zeta^{4} 
                + x^{-2} f_1^2 \zeta^{4} \label{eq:mvsx} 
\eqry
where we have defined $\zeta^{2} = Y_1/Y_T$ for convenience. At 
$x = 1$, precisely at the end point of the physical spectrum,
Eq.(\ref{eq:mvsx}) gives the result $<r^2>_{fit} = 1$.

It is instructive to consider the case in which $\zeta^{2} = 1$ and the
first term in Eq.(\ref{eq:mvsx}) vanishes. (This can occur, for
example, if $\bar{\nu_e}$ is a mass eigenstate so that $\cos\theta_1
=1$ and $\cos\theta_2 = \cos\theta_3 = 0$.)  In this case, no matter
how small the interference is, the fit value of $m^2$ will eventually
become negative far enough away from the end point, assuming all other
quantities were perfectly assigned.  In fact, this could mean that a
value of the mass much less than an $eV$ could affect the fit to the
spectrum several $eV$ below the end point.  As we remarked above when
discussing pseudo-Dirac neutrinos, a very small splitting, as suggested
by the solar neutrino problem, would not appear in this discussion.
Other possible flavor mixings would destroy this special condition.
Existing reports from accelerator measurements~\cite{LSND} and reactor
experiments~\cite{CHOOZ} limit the size of such mixing, but allow it to
be non-zero.

If $\bar{\nu}_e$ is nearly a mass eigenstate, then the second term in
Eq.(\ref{eq:mvsx}) can still dominate, leading to $<r^2>_{fit} < 0$ for
some values of $x$.  The value of $x$ at which $<r^2>_{fit}$ again
becomes positive is a sensitive function of $f_1$ and $\zeta^{4}$.  For
example, taking $\zeta = 0.9996$, which corresponds to a minimum value
of $\sin^2(2\theta_1) \simeq 0.003$ (obtained by setting $\zeta = \cos
\theta_1$, i.e., ignoring possible corrections due to the small, but
generally non-negligible values of the $\varepsilon_k$) in a two flavor
mixing scenario, and $f_1 = 0.075$, which is within the limit on
$X_{S_R}$~\cite{Herczeg}, $<r^2>_{fit}$ reaches $-2.5$ at $x = 45$ and
turns positive a bit beyond $x = 85$.  In such a case, the observed
structure at a given point in the spectrum reflects a mass on a much
smaller scale.

On the other hand, if the quantity $\zeta^{4}$ is small compared to
$1$, $<r^2>_{fit}$ will grow as $x$ increases unless $f_1$ is so large
that $(2 f_1 +f_1^2)\zeta^{4} > (1 - \zeta^{4})$, so that the
derivative with respect to $x$ is negative at $x = 1$.  (In fact, for
$f_1 = 1$, analysis shows that $\zeta^{4}$ must be greater than
$(16/27)$ to get a negative value of $<r^2>_{fit}$ for any value of
$x$.)  For this case to be interesting, a second mass eigenstate must
be involved with the interference term being destructive, so that, far
from the end point, the interference cancels.  Note that, for the cases
at hand, the interference is proportional to the mass so that the more
massive eigenstate can have a smaller admixture.

Since the modern integral
experiments~\cite{TRSK1,TRSK2,TRSK3,MNZ1,MNZ2,MNZ3} provide the best
data available, we defer numerical examples to the next subsection.

\subsection{Fitting fitted integral spectra}  \label{sec-ffis}

To obtain sufficient statistics, experiments that measure differential
spectra make a global fit to data over some range from the endpoint up
to some value of $E_{\nu}$, which translates, in practice, to fitting
over a range of $E_{\beta}$ down to some cut-off value.  This is done
automatically in those experiments that measure an integral spectrum.

For fitting to sums of differential spectra, the fitting procedure,
viewed in terms of theoretical constructs only and ignoring essential
experimental details, corresponds to finding the value of $<r^2>_{fit}$
that minimizes the integral
\beq
I = \int_1^{x_c} dx x^4  \left[Y_1 (1 + \frac{f_1}{x})\sqrt{1-\frac{1}{x^2}} 
- Y_T \sqrt{1-\frac{<r^2>_{fit}}{x^2}} \right]^2  \label{eq:int}
\eeq
where, as in the previous subsection, we have assumed, for simplicity
of presentation, that one mass eigenstate with mass $m_1$ is important,
$x$,$<r^2>_{fit}$, and $f_1$ are as defined previously and $x_c =
E_{\nu}^C/m_1$.

This was the form presented in Ref.(\cite{SG}) and was appropriate for
the earlier differential experiments~\cite{Zurich,LANL,LLNL}.  However,
for the analysis of the integral experiments, one wants to use
Eq.(\ref{eq:apprint}).  The appropriately scaled version is
\beq
N(x_c) = K' Y_1 \left( \frac{1}{3} ({x_c}^2-1)^{3/2}  + \frac{f_1}{2} 
\left[ x_c \sqrt{{x_c}^2-1} - ln{(x_c  + \sqrt{{x_c}^2-1})} \right] \right). 
\label{eq:apintx}
\eeq
Fitting this with the usual SM formula, where $K' Y_T$ is taken from
the fit far from the end point (experimental normalization) and only
one mass eigenstate is included, corresponds to equating
\bqry
N(x_c) & = & K' Y_T \int_1^{x_c} dx x^2 
\sqrt{1-\frac{\rsqft}{x^2}} \NL
& = & K' Y_T \frac{1}{3} \left[ (({x_c}^2 - \rsqft)^{3/2} 
- (1 - \rsqft)^{3/2} \right] \label{eq:smintx}
\eqry
to the same expression in Eq.(\ref{eq:apintx}).  This procedure gives
an  implicit equation for $\rsqft$. In fact the lower limit in
Eq.(\ref{eq:smintx}) makes a very small contribution for most values of
$x_c$ of interest, so that, to a good approximation, one may drop that
term and obtain an explicit equation for $\rsqft$.

We used that equation to search through parameter space to find sets
that gave a reasonable representation of the latest Mainz
data~\cite{MNZ3}.\footnote{The experimentalists report substantial
consistency of their results~\cite{TRSK3,MNZ3}; we have simply found it
more convenient to make use of the Mainz presentation of their data.}
We emphasize again that we have not constructed a fit, even to the
representation in terms of $\msqft$, but rather have selected a set of
parameters that give a reasonable representation of the published
results. We have done this to demonstrate that the inclusion of
possible interference effects of non-SM currents shows no obvious
evidence of being at odds with the data.

To that end we offer a comparison in Figs.\ 3-6.  The data points were
read from a reproduction of Fig.\ 3 of a preprint of Ref.(\cite{MNZ3})
and should not be taken as a precisely accurate representation. In our
Fig.\ 3, open circles correspond to  the Q2 data set of their Fig.\ 3;
in our Fig.\ 4, open squares correspond to their Q3 data set; in our
Fig.\ 5, open diamonds correspond to their Q4 data set and in our
Fig.\ 6, asterisks correspond to their Q5 data set.  No attempt was
made to generate fits; rather the parameter sets were chosen to give a
maximum negative $\msqft$ about $30 eV$ below the endpoint, $\fep$, and
to turn $\msqft$ positive at $E_{\beta} \approx 18350 eV$.  In the
figures, the various calculated curves are labelled by the value of
$f_1$.  In Table~I, we list the full parameter set for each curve,
where the parameters are $f_1$, $m_1$ and $\zeta^{2}$.  The final 
row, labelled $\sin^2 2\theta$, is the minimum value one would
deduce for that quantity in a two component oscillation formula,
obtained by taking the value of $\zeta^{2}$ to equal $\cos^2\theta_1$,
i.e., treating the $\varepsilon_k$ as negligibly small.

\begin{center}
\parbox{6in}{Table I. Interference parameters for theoretical
curves plotted in Figs.\ 3-6.}
\begin{tabular}{||c|c|c|c|c|c||}       \hline
\setlength{\arraycolsep}{1.0in}
$~f_1~$    &~~~.05~~~&~~~.07~~~&~~~.09~~~&~~~.11~~~&~~~.13~~~ \\ \hline
$m_1 (eV)$ &  1.603  &  2.228  &  2.888  &  3.431  &  4.334   \\ \hline
$\zeta^{2}$&  .99954 & .99910  &  .99850 &  .99783 &  .99675  \\ \hline
$\sin^2 2\theta$ & .00185 & .00360 & .00600 & .00870 & .01300 \\ \hline
\end{tabular}
\end{center}

Since the authors of \cite{MNZ3} chose not to combine these data sets,
for reasons they describe, we also do not attempt to combine them.  In
fact, as we discuss in the next section, there are theoretical reasons
to avoid combining runs taken at different times.  In spite of that
restriction, we believe that Figs.\ 3-6 make the point that, for values
of the parameters which are within reason, the possibility of
interference between the SM current and a weaker, non-SM current can
give the observed negative values of $\msqft$ when the data is analyzed
under the assumption that only the SM current is present.

\subsection{Discussion} \label{sec:ftdisc}

As we showed in Sec.(\ref{sec-ffis}), it is possible to generate values
of $\msqft$ that resemble the published data for values of $f_1$
between $.05$ and $.13$.  The lower end of this range is easily
accommodated by existing limits, even without invoking the possibility,
discussed in Sec.(\ref{sec-limits}) that these limits are only
appropriate for a sum over the mass eigenstates.  The upper end of the
range may be accommodated by invoking this last point or by noting that
more than one non-SM current may be affecting the Tritium data in
combination.

While we have not carried out an exhaustive analysis of the allowed
parameter space, it is worth noting that, by studying
Eq.(\ref{eq:mvsx}), it is possible to get some sense for the allowed
parameter ranges.  For example, requiring $\msqft = -10 eV^2$ at
$E_{\nu} = 100 eV$ relates $m_1$ to $f_1$. In particular,  if $f_1 <
0.11$, then $1 eV < m_1 < 10 eV$.  The lower limit comes from the
obvious effect that the strength is $m_1 f_1$ and the upper limit from
the fact that the kinematic effect of $q_{\nu}$ begins to close the
phase space.  We do not mean to emphasize $f_1 = 0.11$ particularly;
our point is that the present results strongly suggest that fitting to
data using our functional form will give $m_1$ near this approximate
range.

The required mass eigenvalues, on the other hand, would appear to be in
conflict with recent limits from double beta decay~\cite{HbrgMscw}.
However, those limits are on the expectation of the Majorana mass and,
as Wolfenstein pointed out~\cite{LW2}, for a pseudo-Dirac neutrino it
is only the difference between the mass eigenstates that is limited by
bounds from neutrinoless double beta decay experiments.  Taken together
with the tiny $\Delta m^2$ inferred from solar neutrino
studies~\cite{solnsum}, the values of $m_1$ (in the few $eV$ range)
that appear in this analysis strongly imply that the electron neutrino
is either a Dirac particle or a pseudo-Dirac particle.~\cite{GSMK,KC}
The fact that $\theta_1$ is very small (inferred from the small value
of $1 - \zeta^{2}$) in all of the preferred parameter sets is
consistent with a small amount of flavor mixing, but the fact that it
cannot be zero also indicates that some flavor mixing is required.

We note here that, if this scenario obtains, then there do exist
additional neutrino degrees of freedom beyond those active in SM
interactions.  Such degrees of freedom are usually termed ``sterile'';
however, our analysis of Tritium beta decay clearly implies that these
``sterile'' neutrinos necessarily participate in new interactions
beyond the SM.  In Sec (\ref{sec:nceff}) we discuss other implications
of this observation.

\section{Environmental Effects}

We~\cite{Clouds} have previously studied the consequences that may
arise if there is a very light scalar particle coupled only to
neutrinos.\footnote{X.-G.\ He {\it et. al.}~\cite{He} have shown that
such a scenario is possible in extended gauge theories.} The primary
effect is that, wherever in space the classical scalar field arising
from such interactions is non-zero, the effective mass of the neutrino
is altered from its vacuum mass (the mass a neutrino would have in a
region of space devoid of other neutrinos, but including all field
theoretic contributions).  It is this effective mass at the site of a
Tritium beta decay event which will govern the interference effects
discussed above.

A second result reported in Ref.(\cite{Clouds}) is that, for a wide
range of parameters which are not in conflict with known data,
neutrinos will form clouds during an early stage in the development of
the Universe.  The extent and density of such clouds will depend on the
details of the neutrino vacuum masses, the mass of the scalar and the
size of the coupling between the scalar and neutrinos. However, there
is no known impediment to considering a length scale between a solar
radius and the size of a solar system, with the range of the scalar
field being somewhat smaller.  If the scale of the particular cloud in
which the Solar system developed is on the order of the Earth's orbit,
it would be possible that the strength of the scalar field sampled at
the Earth would vary with the Earth's position along its orbit, leading
to a time dependence of the effective mass.

Alternatively, the cloud itself could be undergoing various forms of
collective motion (rotations of an ellipsoidal shape or vibrations in
various modes), which will be reflected in variations of the strength
of the scalar field.

The exact form of the variation of the effective mass as observed on
Earth would further depend on unknown details.  For example, if the
electron anti-neutrino is mostly aligned with the heaviest vacuum mass
eigenstate, then the stronger the scalar field, the smaller the
effective mass.  If, however, as we showed in Ref.(\cite{MSWlike}), the
electron antineutrino is mostly aligned with one of the lighter vacuum
mass eigenstates, as the scalar field gets larger, the magnitude of the
effective mass increases. What is uniformly true is that, as the scalar
field varies, the effective mass varies.  Further, if the effective
mass is a small fraction of the vacuum mass, small percentage
variations of the scalar field can lead to larger percentage variations
of the effective neutrino mass.

Another consequence that follows from this conjecture, as discussed in
detail in~\cite{Clouds}, is that neutrinos could not constitute a hot
dark matter component of the Universe during the formation of large
scale structures, since they would be ``locked up'' within massive
clouds with non-relativistic net kinetic energies.  Thus a number of
the reported limits on neutrino mass from cosmology would need to be
revisited.

While there is, of course, no proof that such a background scalar field
exists in the region of the Earth (or at all), the fact that it cannot
be ruled out suggests that there is a value to independently analyzing
experiments done at different times.  If there is a time dependence in
the effective mass matrix, this may imply a time dependence to the
resulting direction cosines.  In the formulation presented here, that
would affect both $\cos\theta_1$ and the value of $f_1$ through this
and its dependence on $\cos\hat{\theta}_1$.  Hence all the parameters
of the fit may demonstrate a time dependence, and the correlations are
very hard to predict {\it a priori}.  As bizarre as this possibility
may seem, it is clearly prudent to await further experimental
information before discarding it.

\section{Neutrino endpoint effects} \label{sec-neutep}

The interference terms discussed above display a general symmetry
between the $\beta$ and the neutrino.  M. Goldhaber~\cite{Maurice} has
raised the question of possible observable effects at the opposite end
of the spectrum, where the neutrino carries away all of the available
energy.  In fact, the only terms that would be observable are
complimentary to those discussed above, since any terms proportional to
the neutrino mass eigenvalues will be completely dominated by
$E_{\nu}$.  This leaves the term proportional to $\rho_{S_R}
\frac{m_e}{E_{\beta}}$.  Since the scale of variation with energy is
set by $m_e$, we would expect the variation to occur over several
$MeV$.  Consequently, Tritium is not the place to look for such
effects.

As the Fermi function varies rapidly at small $E_{\beta}$, the analysis
must make use of the full form given above.  Furthermore, given the
enhancement of the very low energy electron spectrum, compared with the
suppression for a positron spectrum, electron emitters may be
preferred. This particular current is the most severely constrained,
and this part of the spectrum corresponds to large $E_{\nu}$ so that
threshold effects are not relevant. Nonetheless, the fact that the
scale of the variation is set by $m_e$ may make it possible to see such
an effect in neutrino spectra of future experiments.

Although the observation of interference effects in such a transition
would not directly impact the interpretation of Tritium beta decay, it
would demonstrate the existence of non-SM currents and would be a very
interesting result in its own right.

\section{Implications for neutral currents} \label{sec:nceff}

As we discussed in section (\ref{sec:ftdisc}), the combination of our
analysis of Tritium beta decay and other experimental results strongly
suggests that the electron neutrino is a pseudo-Dirac object with new
interaction(s) involving the components which are sterile in the SM.
If these new interactions devolve from Higgs mediated scalar currents,
as in supersymmetric extensions of the SM~\cite{susy}, or from
Right-chiral Vector currents, as in Left-Right symmetric
models~\cite{lrsym}, there will be associated new neutral current
interactions for the ``sterile'' components of the neutrino fields.  In
fact, such neutral currents occur in most proposed extensions of the
SM.

These new interactions could affect the interpretation of any
experiment sensitive to neutral currents.  For example, in the case of
the Sudbury Neutrino Observatory experiment (SNO)~\cite{SNO},
oscillation of solar neutrinos into a ``sterile'' component could
produce a neutral current signal intermediate in strength between that
expected for oscillations among active neutrino components only and the
reduction of signal observed for the charged current interactions.
Absent any special quantum coherence effects, the bounds on the
strength of any new charged current interactions suggest that the
additional neutral current signal provided by this mechanism is not
likely to be sufficiently large to confuse an oscillation into
``sterile'' components with those among active neutrinos of different
flavors.  Nonetheless, the SNO experimental group must specify the
bounds on ``sterile'' neutral current strengths assumed in the analysis
of their data.

Similarly, if, as one expects, these considerations apply to other
flavors as well, experimentalists such as those involved in the
SuperKamiokande experiment~\cite{superK} need to consider the possible
strength of neutral current interactions of ``sterile'' neutrinos.
That particular case is exacerbated by the fact that the effect of
interest is proportional to forward scattering, hence depends on
amplitudes rather than rates.

\section{Conclusions}

In this paper we have examined the possibility that interference
between the SM Left-chiral current and a weaker, non-SM current with a
different Lorentz character may be the origin of the ``anomaly'' in the
Tritium beta decay spectrum near the end point. On general theoretical
grounds, it is expected that such currents must exist, the only
uncertainty being their strength. To avoid the unwarranted prejudice
that only the SM interaction contributes to the decay, experimentalists
should include these interference terms when fitting data.

Given the small variation in the electron energy, $E_{\beta}$, we have
presented formulas for differential spectra, Eq.(\ref{eq:apprdif}), and
for integral spectra, Eq.(\ref{eq:apprint}), which are good
approximations appropriate for fitting to data. At a minimum, these
should be employed to obtain reasonable parameter values from which to
initiate searches with more complete expressions.

Explicitly, we recommend the use of Eq.(\ref{eq:apprint}), with $k=1$
and $\phi_1 = f_1 m_1$, for the analysis of ongoing integral
experiments~\cite{TRSK3,MNZ3}, with all of sums over molecular end
points implied by Eqs.(\ref{eq:enchila} and \ref{eq:finaldif}) and
experimental details included. This analysis should be carried out
independently of other experimentally derived constraints as those
generally reflect differing parameter combinations. In particular, this
experiment is uniquely affected by neutrino mass thresholds.

Using a characterization of the data in terms of $\msqft$, we have
shown that our interpretation of the negative values reported to date
is possible for parameter values which are not, in fact, in conflict
with other experimental constraints.  For consistency with experimental
results on neutrino oscillations and neutrinoless double beta decay,
the electron neutrino must be either a Dirac or a pseudo-Dirac object
with a small CKM-like mixture to other flavor eigenstates.

Finally, we examined three other related considerations: One is the
possibility that neutrino endpoint effects might appear to be time
dependent due to environmental factors. The second is the complementary
effect of new interactions, as discussed here, on the high neutrino
energy (low electron energy) end of beta spectra~\cite{Maurice}.
Lastly, and perhaps most dramatically, we noted that our analysis, when
combined with other results, strongly implies the existence of new
interactions involving ``sterile'' neutrinos. This last, in turn, has
important implications for the interpretation of experiments studying
neutral current interactions.

\section{Acknowledgments}

We are happy to acknowledge valuable conversations on these topics with
Maurice Goldhaber, Bill Louis and Peter Herczeg.  We thank Christian
Weinheimer for careful reading of and comments on drafts of this paper.
This research is partially supported by the Department of Energy under
contract W-7405-ENG-36, by the National Science Foundation and by the
Australian Research Council.  One of us (GJS) acknowledges the
hospitality of the Institute for Nuclear Theory at the University of
Washington on several occasions, where portions of this work were
carried out.

\newpage

\section{Figure captions}

Figure 1.  Current eigenstate neutrinos displayed in the space spanned
by the three mass eigenstates.  The vectors denoted by $\nu_f$ are
current eigenstates for the SM $SU(2)_L$ current, those denoted by $
\hat{\nu}_f $ are current eigenstates for whatever other charged
current one is considering.

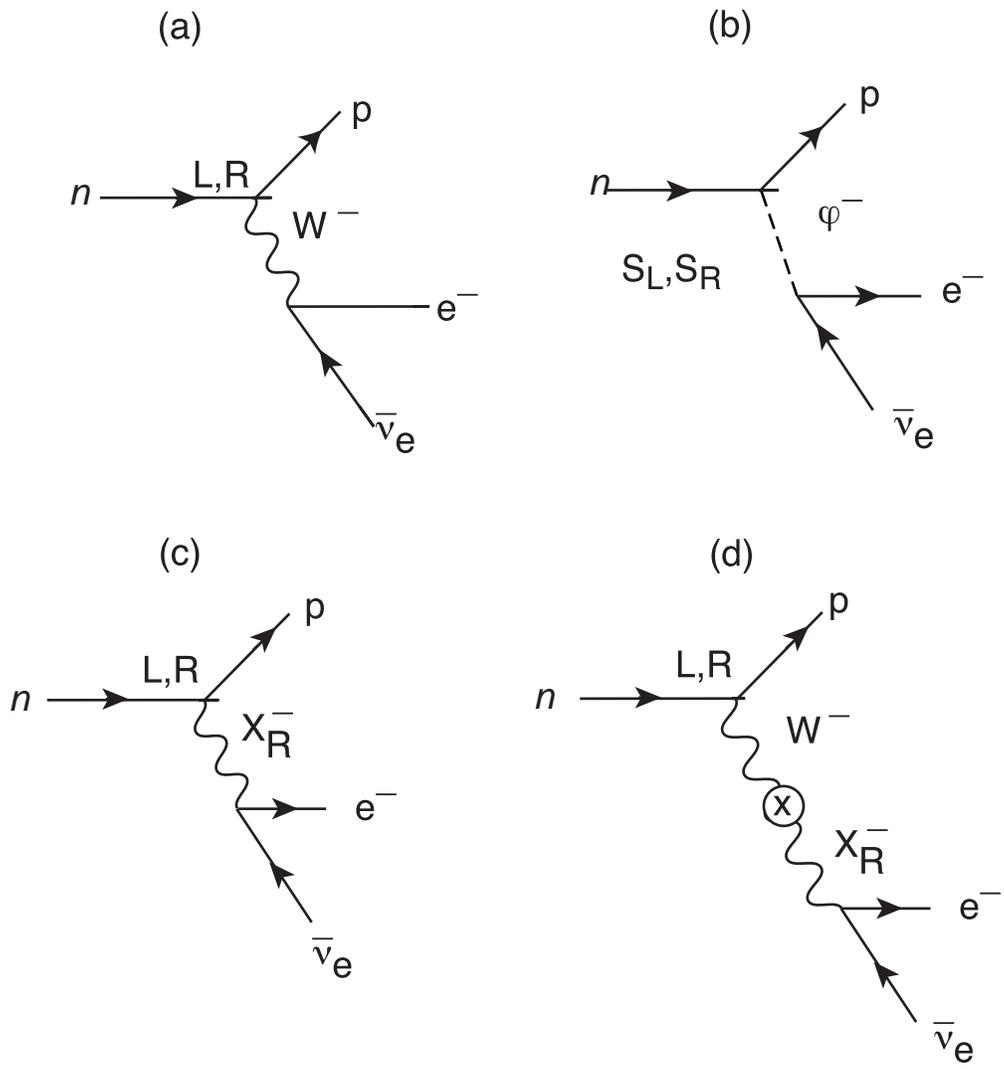
Figure 2.  Diagrams contributing to nuclear beta decay under various
scenarios.  a) SM interaction.  Hadronic renormalization produces both
L and R hadronic currents in the effective Hamiltonian.  b)  Scalar
exchange.  Recoil effects suppress the hadronic Pseudo-scalar
coupling.  In principle the two leptonic couplings, $S_L$ and $S_R$ can
be different.  c) Direct Right-chiral Vector couplings.  d)  Possible
mixing between the  $X_R^{-}$ and $W^{-}$.  The analogous diagram in
which the $W^{-}$ couples to the lepton current gives no observable
change in the Tritium spectrum.  Again, hadronic renormalization will
lead to both L and R effective hadronic currents in both diagrams c)
and d).

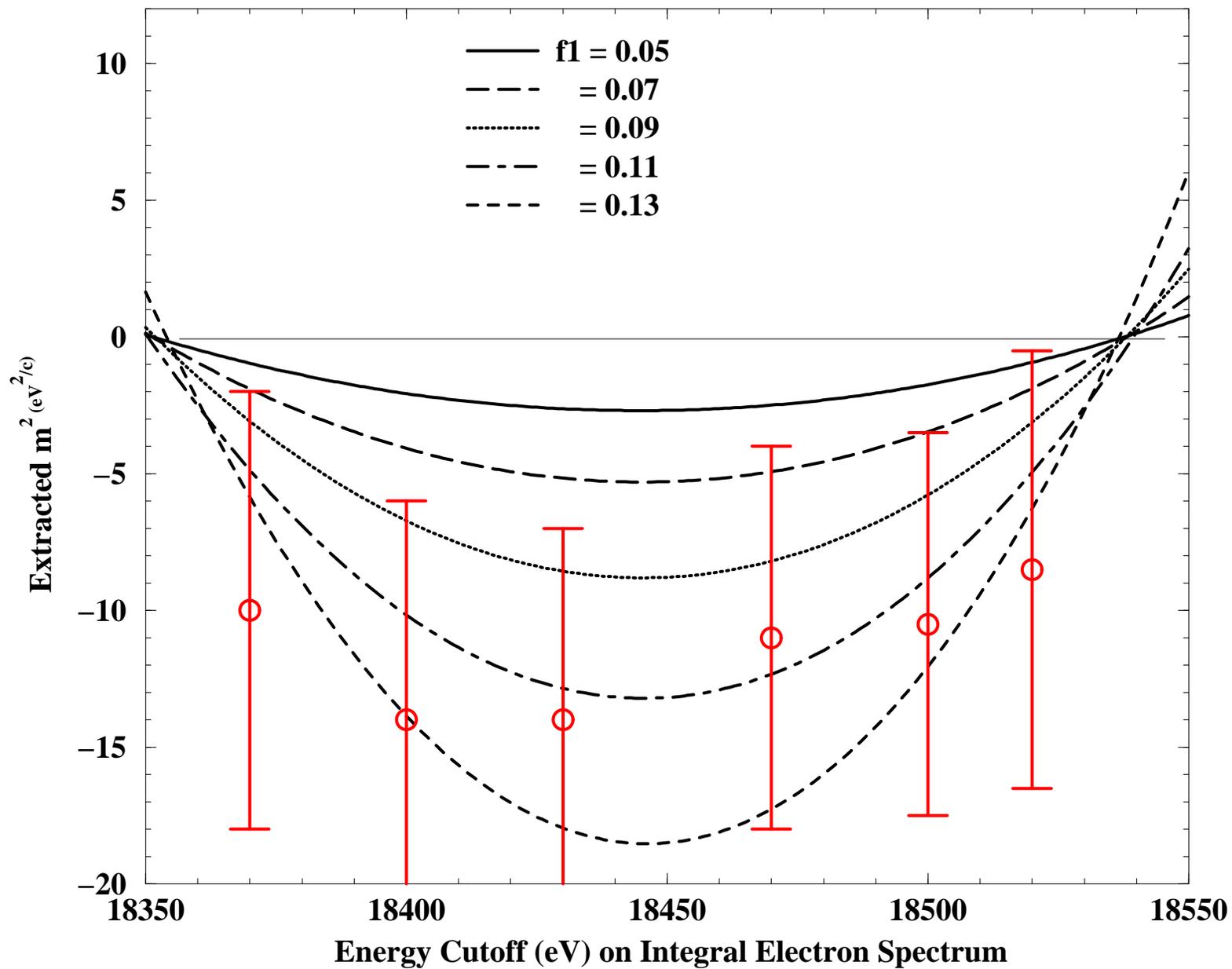
Figure 3. Neutrino mass-squared extracted from Mainz data set Q2 vs.
integral cutoff on electron energy and corresponding results from the
SM model analysis of a spectrum including interference effects, for
various parameter values. (Details in text.)

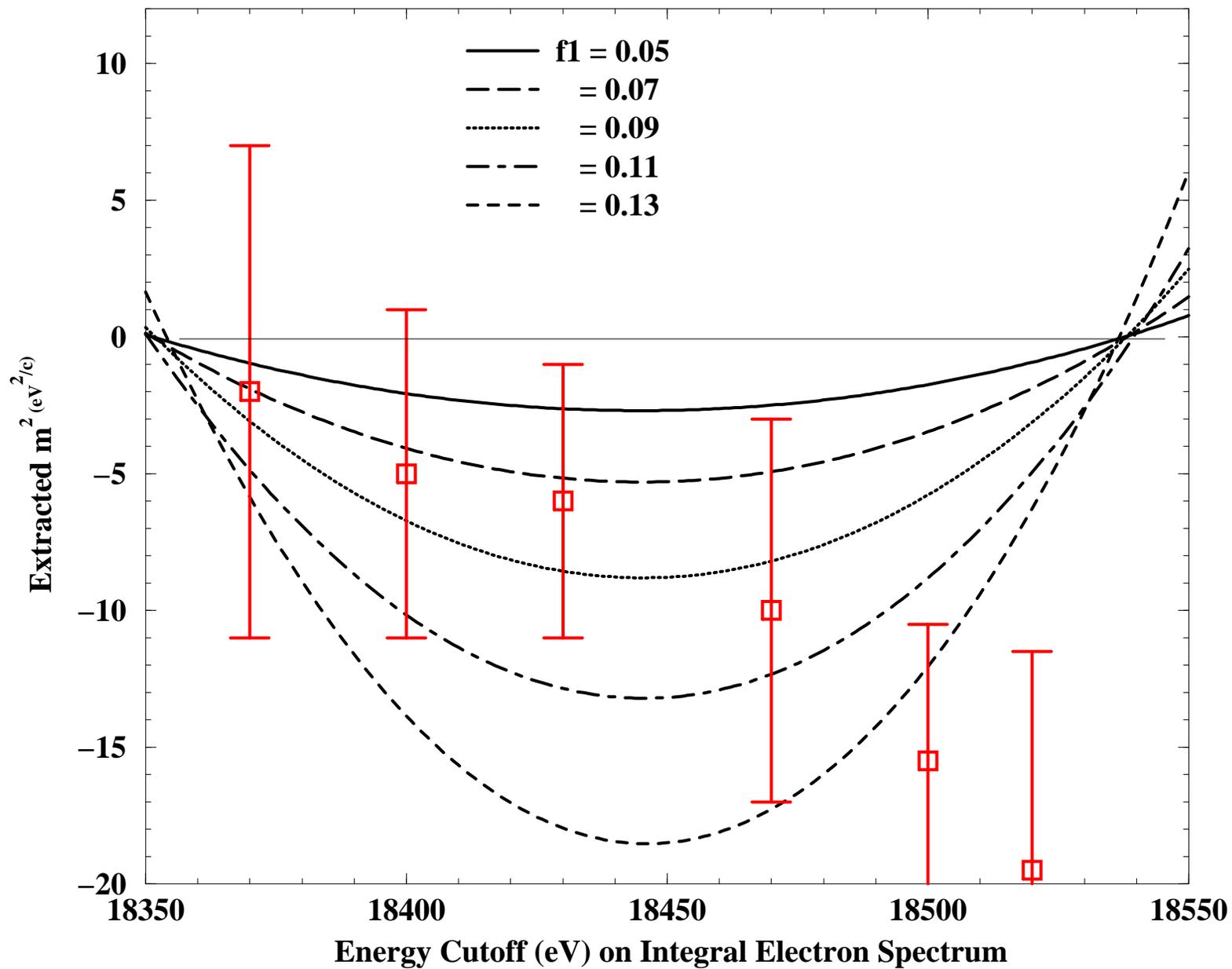
Figure 4. The same as Fig.3 for Mainz data set Q3.

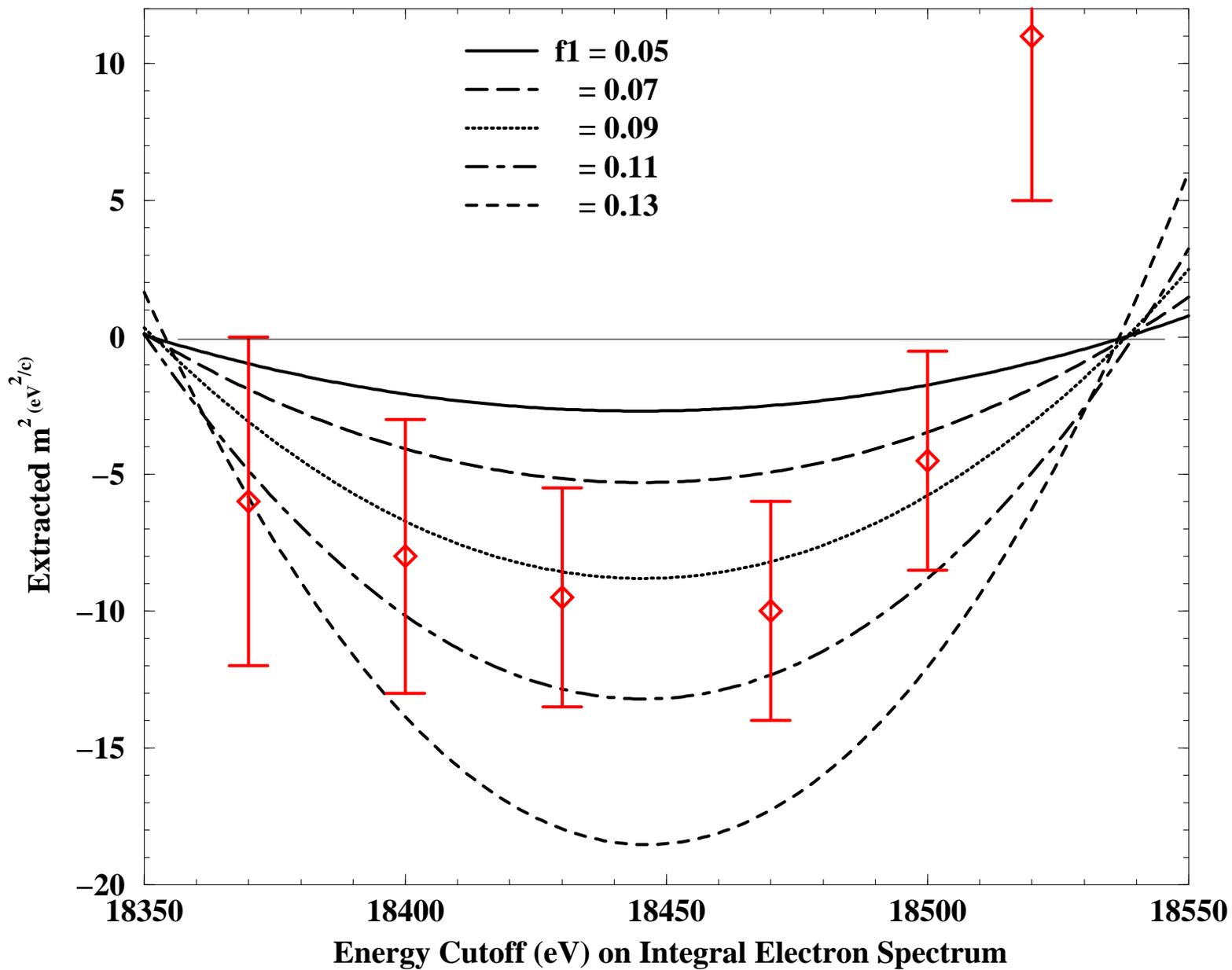
Figure 5. The same as Fig.3 for Mainz data set Q4.

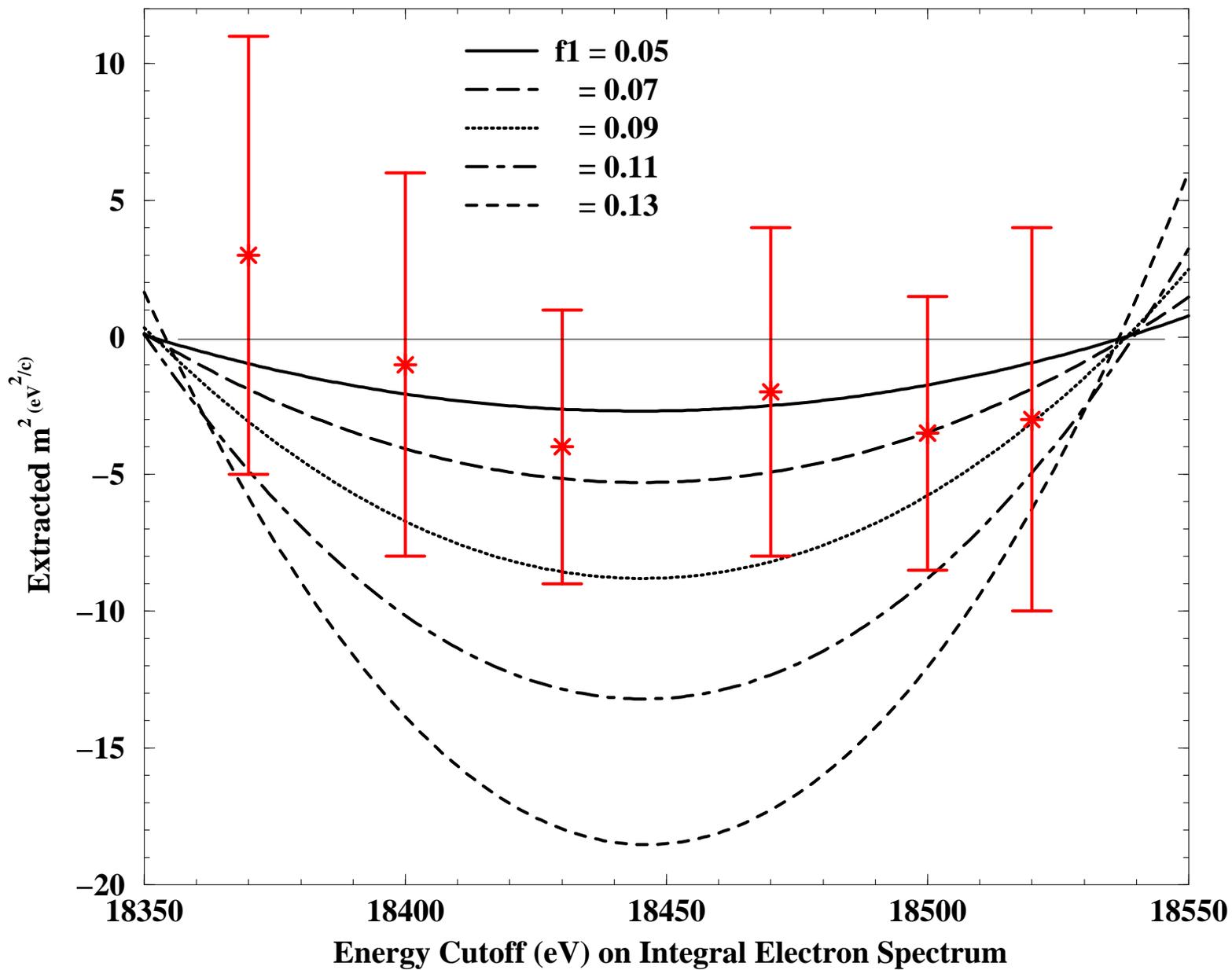
Figure 6. The same as Fig.3 for Mainz data set Q5.

\end{document}